\documentclass[twoside,slac_one]{revtex4}
\usepackage{graphicx}
\usepackage{fancyhdr}
\usepackage{amsmath} 
\usepackage{bm}
\usepackage{amsxtra}
\usepackage{amssymb}
\usepackage{amsthm}
\usepackage{latexsym}
\usepackage{lscape}
\usepackage{atlasphysics}

\usepackage{multirow}
\usepackage{subfigure}
\usepackage{mathrsfs}

\newcommand{\ptmiss}{{\vec{P}}_\mathrm{T}^\mathrm{\; miss}}

\newcommand{\ph}{\phantom{0}}

\begin{document}

\title{Interpretations of SUSY Searches in ATLAS with Simplified Models}

%

\author{Hideki Okawa, on behalf of the ATLAS Collaboration}
\affiliation{Department of Physics and Astronomy, University of California at Irvine, Irvine, CA, USA}

\begin{abstract}

We present the status of interpretations of Supersymmetry (SUSY) searches 
in ATLAS at the Large Hadron Collider (LHC) using simplified models. Such models allow a systematic scan 
through the phase space in the sparticle mass plane, and in the corresponding 
final state kinematics. Models at various levels of simplification have been 
studied in ATLAS. 
The results can be extrapolated to more general new physics 
models which lead to the same event topology with similar mass hierarchies.
Searches in the no-lepton channel with 1.04~\ifb of data from 2011 and 
the same-sign dilepton channel with 35~\ipb of data from 2010 are presented. 
No excess above the Standard Model expectation is observed, and the
results are interpreted using the simplified models.
\end{abstract}

\maketitle

\thispagestyle{fancy}


\section{Introduction}

ATLAS~\cite{ATLAS_exp} at the Large Hadron Collider (LHC)~\cite{LHC}
recorded 2.55 \ifb of data by the end of 
August~\cite{pub_lumi}. 
We are truly entering the TeV-scale, and starting to have a wide
coverage over possible Supersymmetry (SUSY) models. In order to 
ensure that all the relevant phase space is covered, 
searches should not be over-driven by specific high energy scale SUSY models. 

The simplified model approach~\cite{marmo}\cite{alwall}\cite{alves} is one of 
the most promising model-independent strategies for new physics searches,
being widely discussed both in the experimental and theoretical 
communities~\cite{lhcnewphys}. 

In these proceedings, the overview of the simplified model
approaches used in ATLAS is presented. Brief descriptions of the ATLAS
detector and dataset used in the analyses follow. 
Searches and simplified model interpretations 
in the no-lepton channel with 1.04~\ifb of data from 2011 and 
the same-sign dilepton channel with 35~\ipb of data from 2010 are described.   
The status of b-jet channel using simplified models is mentioned 
in~\cite{bart}. 

\section{Simplified Models}

Simplified models are effective models built
with the minimal particle content necessary to produce
SUSY-like 
final states contributing to the channels of 
interest 
and are parametrized directly in terms of the sparticle masses.
They naturally allow for reducing the dimensionality of the
theoretical parameter space to two to four sparticle mass
parameters and relevant branching ratios.
Figure~\ref{fig:simpleDecays} shows the simplified decay chains 
considered in ATLAS.  
This approach allows to scan the whole sparticle mass plane without 
imposing a strict relation on the gaugino masses as are the cases 
in many of the high energy scale SUSY models such as 
mSUGRA~\cite{msugra}\cite{cmssm}. 
Furthermore, the results can be expressed in terms of limits on cross-section 
times branching ratios as a function of new particle masses,
separately for each event topology, or even down to each diagram, thus
disentangling assumptions on the relative couplings at each vertex. 
The simplified model framework is a complementary approach to 
phenomenological Minimal Supersymmetric Standard Model (pMSSM)~\cite{rizzo}.

The results are therefore generic and can serve as an interface 
to theories with additional Standard Model (SM) partner particles 
({\it e.g.} Universal Extra Dimensions, UED~\cite{UED}), which make predictions 
in these topologies.

\begin{figure}[t]
\centering
\includegraphics[width=80mm]{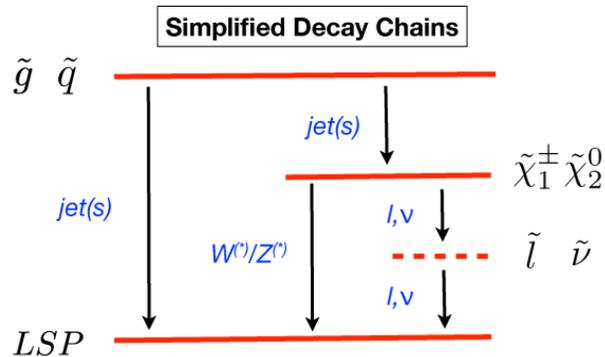}
\caption{Simplified decay chains considered in ATLAS.} 
\label{fig:simpleDecays}
\end{figure}

\section{The ATLAS Detector and Data Samples}
\label{Sec:Det}

The ATLAS detector is a multipurpose particle physics 
apparatus with a forward-backward symmetric cylindrical geometry and 
near 4$\pi$ coverage in solid angle~\footnote{ATLAS uses a right-handed 
coordinate system with its origin at the nominal interaction point (IP) 
in the center of the
detector and the $z$-axis coinciding with the axis of the beam pipe. 
The $x$-axis points from the IP 
to the center of the LHC ring, and the $y$ axis points upward. 
Cylindrical coordinates ($r$,$\phi$) are used in the transverse plane, 
$\phi$ being the azimuthal angle around
the beam pipe. The pseudorapidity is defined in terms of the polar angle 
$\theta$ as $\eta = -{\rm ln} \tan(\theta/2)$.}. 
The inner tracking detector (ID) covers the pseudorapidity range $|\eta|<2.5$, 
and consists of a silicon pixel detector, 
a silicon microstrip detector (SCT), and, for  $|\eta|<2.0$, a transition 
radiation tracker (TRT).
The ID is surrounded by a thin superconducting solenoid providing a 2T 
magnetic field.
A high-granularity liquid-argon (LAr) sampling electromagnetic calorimeter 
covers the region $|\eta|<3.2$. 
An iron-scintillator tile hadron calorimeter provides coverage in the 
central rapidity range of $|\eta|<1.7$. 
The end-cap and forward regions, spanning $1.5<|\eta|<4.9$, are instrumented 
with LAr calorimetry for both 
electromagnetic and hadronic measurements. 
The muon spectrometer (MS) surrounds the calorimeters and consists of three 
large air-core superconducting toroids, 
a system of precision tracking chambers up to $|\eta|<2.7$, and detectors for 
triggering in the region of $|\eta|<2.4$.

For the no-lepton search, the data are collected in 2011 with the LHC operating 
at a centre-of-mass energy of 7~\TeV.
Application of beam, detector and data-quality requirements results in
a total integrated luminosity of $1.04 \pm 0.04$~fb$^{-1}$
with an estimated uncertainty of 3.7\%~\cite{lumi2011}.
The trigger requires events to contain a leading jet with a transverse momentum 
($\pt$), measured at the raw electromagnetic (EM) scale, above $75 \GeV$ and 
missing transverse energy (\met) above $45 \GeV$.  

The data used for the same-sign dilepton analysis are recorded in 2010 at the LHC at 
a center-of-mass energy of 7 TeV. 
Similarly to the 2011 data, the application of beam, detector, and data-quality 
requirements results in a total 
integrated luminosity of 35~\ipb with an estimated uncertainty of 
11\%~\cite{lumi}.
The data are selected with single lepton (electron or muon) triggers. 
The detailed trigger requirements vary throughout the data-taking period 
owing to the rapidly increasing LHC luminosity and the commissioning of 
the trigger system, but always ensure that leptons with 
\pt$>$20~GeV lie in the efficiency plateau. 
Monte Carlo (MC) simulation samples for background processes are 
generated as described in~\cite{2lep-paper}. 

\section{Object Definition}
\label{sec:obj}

Jet candidates are reconstructed using the anti-$k_t$ jet clustering 
algorithm~\cite{antiKt}\cite{antiKt2} with a distance parameter of $0.4$. 
The inputs to this algorithm are three-dimensional clusters of calorimeter 
cells~\cite{calocluster} seeded by those with energy significantly above
the measured noise. Jet momenta are constructed by performing a
four-vector sum over these calorimeter clusters, treating each as an
$(E,\vec{p})$ four-vector with zero mass.  These jets are corrected for
the effects of calorimeter non-compensation and inhomogeneities by
using $\pt$ and $\eta$-dependent calibration factors
based on Monte Carlo (MC) simulation and validated with extensive
test-beam and collision-data studies \cite{atlas-jetcal}. 
Furthermore, for 2011 data, the average additional energy due to multiple 
events in a single beam crossing (pile-up) is subtracted using correction 
constants extracted from data, and the reconstructed jet is modified such 
that the jet direction points to the primary vertex of the interaction 
instead of the geometrical center of the ATLAS detector. 
Only jet candidates with transverse momenta $p_T > 20$ \GeV{}
are subsequently retained.

Electron candidates are required to have $\pt > 20$ GeV, $|\eta| < 2.47$, 
and pass the `medium' electron shower shape and 
track selection criteria of~\cite{atlas-electron2011}.
Muon candidates \cite{atlas-muon2011} are required to 
have $\pt > 10$ GeV (and 20 GeV for the same-sign 2-lepton analysis) 
and $|\eta| < 2.4$.  

The measurement of the missing transverse momentum $\ptmiss$ and 
its magnitude, missing transverse energy $\met$ is based on 
the transverse momenta of all electron and muon candidates, all jets which are not also electron candidates, and all
calorimeter clusters with $|\eta|<4.5$ not associated to those objects.  Since tau-lepton candidates are not used
in these analyses, the term ``lepton'' will refer only to electrons and muons.

Following the steps above, overlaps between candidate jets with $|\eta|<2.8$ and leptons are resolved using the method of~\cite{ATLAS_detPerf} as follows.
First, any jet candidates lying within a distance $\Delta R = \sqrt{(\Delta \eta)^2 +
(\Delta \phi)^2}
=0.2$ of an electron are discarded,
and then any electron or muon candidates remaining within a distance
$\Delta R =0.4$ of any surviving jet candidate are discarded. 
Next, all jet candidates with $|\eta|>2.8$ are discarded. 
Finally, the remaining electron, muon and jet candidates are considered
``reconstructed'', and the term ``candidate'' is dropped.

\section{No-lepton Search}

The SUSY search with all-hadronic and \met ~signature
(so called ``No-lepton Mode'') is considered to be one of 
the ``golden channels.'' It has the largest coverage over
possible phase space of R-parity conserving pMSSM 
models~\cite{rizzo}.

In this search, three simplified models are considered as shown 
in Figure~\ref{fig:simplifiedModel_nolep}, where 
gluinos and light-flavor squarks are initially pair-produced. 
Direct decays to the lightest supersymmetric particles (LSPs)
are considered, which are the simplest decays contributing to 
the no-lepton channel~\cite{alwall}. 

\begin{figure} [t]
\centering
\begin{tabular}{ccc}
\includegraphics[width=0.3\linewidth]{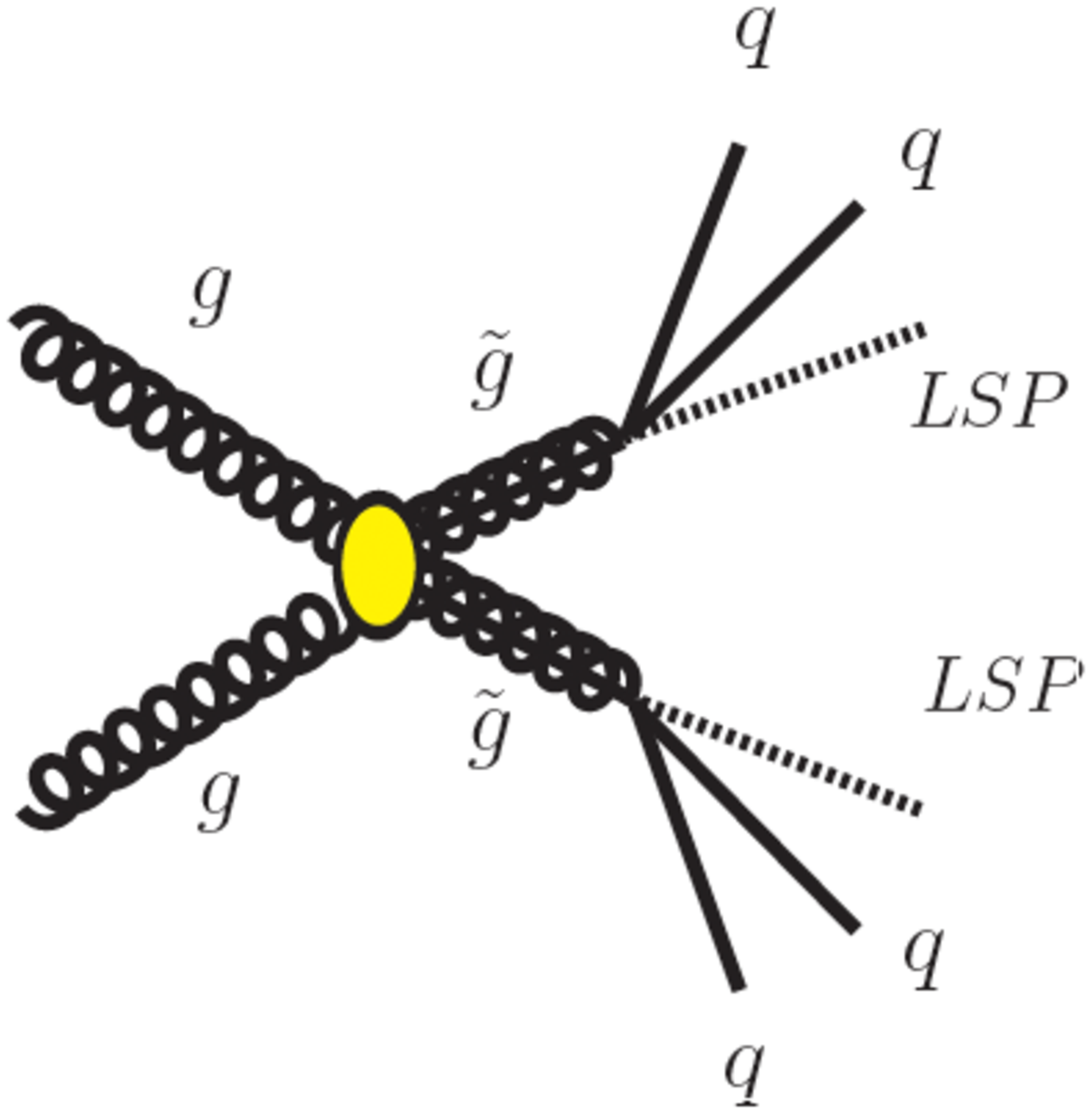}
\includegraphics[width=0.3\linewidth]{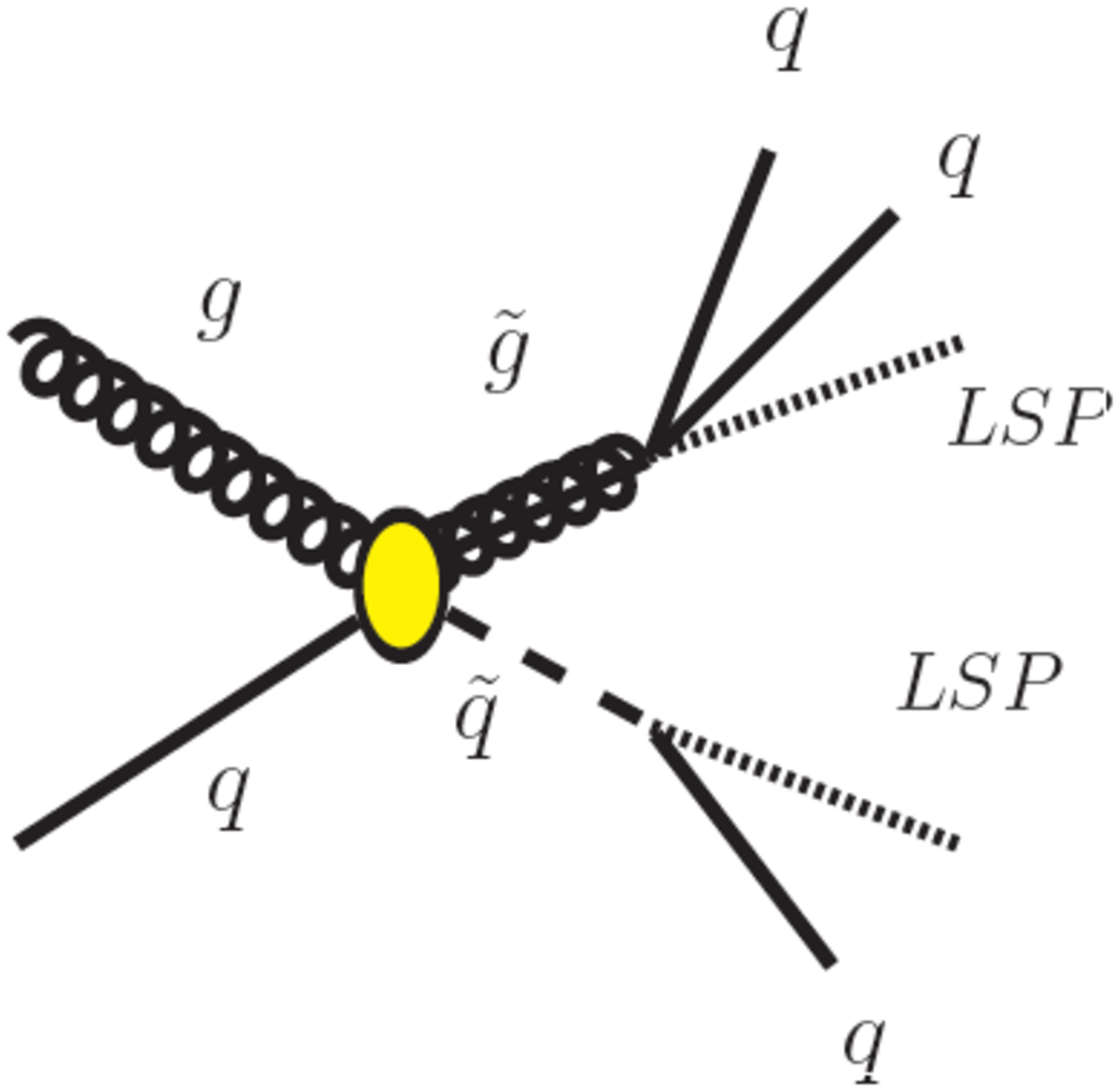} 
\includegraphics[width=0.3\linewidth]{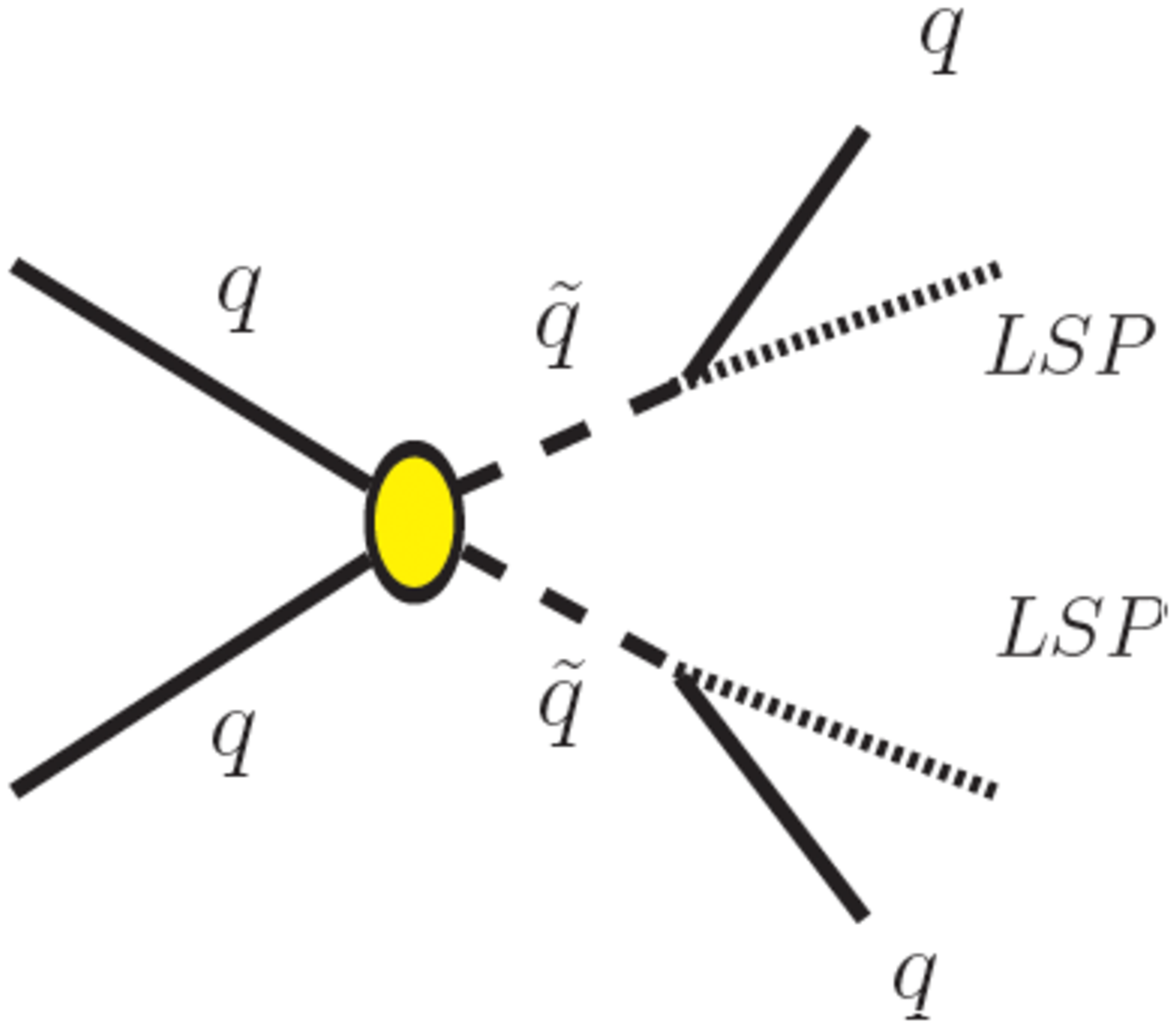}
\end{tabular}
\caption{Simplified models considered for the no-lepton channel, where 
gluinos and squarks directly decay down to the LSP in the gluino-gluino (left),
gluino-squark (center), and squark-squark (right) processes.}
\label{fig:simplifiedModel_nolep}
\end{figure}

Five signal regions are considered to cover various regions of the
gluino-squark mass plane as shown in Table~\ref{tab:0lep_SR}.
First of all, events are discarded if electrons and muons with $\pt > 20$ GeV
remain. In order to suppress detector noise and non-collision 
background, events are required to have the primary vertex 
associated with at least five tracks, and are discarded if 
any jets fail quality selection criteria described in ~\cite{jetcleaning}.
Cuts on \met ~and the leading jet $p_T$ are applied to be on the efficiency 
plateau of the trigger requirement. Further selections on the sub-leading
jets are considered to cover event topologies from $\tilde{q}\tilde{q}$, 
$\tilde{g}\tilde{q}$, and $\tilde{g}\tilde{g}$ processes, where
the final states consist of at least two, three, and four jets respectively.  
Effective mass ($\meff$) is defined as the scalar sum of \met and the $\pt$
of the two, three, four leading, or all the jets with $\pt>40$~GeV used for 
each signal region.
It is known to be approximately proportional to the mass of initially produced 
sparticles~\cite{Meff}. 
$\Delta \phi(\textrm{jet},\ptmiss)_\mathrm{min}$ and $\met/\meff$ cuts are 
applied to suppress the QCD background. Two signal regions are considered
for the four-jet case using different requirements on $\meff$ to cover 
smaller and larger mass splittings between the sparticles.

\begin{table}[t]
  \begin{center}\renewcommand\arraystretch{1.2}
\renewcommand\tabcolsep{1.2mm}
    \begin{tabular}{| l l | c|c|c|c|}
 \hline \hline
 & Signal Region &$\geq$ 2-jet & $\geq$ 3-jet & $\geq$ 4-jet & High mass \\
\hline
\hline
& $\met$            & $>130$ &  $>130$ &  $>130$  & $>130$ \\ 
& Leading jet $\pt$       & $>130$ &  $>130$ &  $>130$  & $>130$ \\ 
\hline
& Second jet      $\pt$   & $>40$ &  $>40$ &  $>40$ & $>80$\\ 
& Third jet      $\pt$       &  --         &  $>40$ &  $>40$ & $>80$\\ 
& Fourth jet      $\pt$       &  --        &     --       &  $>40$ & $>80$\\ 
\hline
&$ \Delta \phi(\textrm{jet},\ptmiss)_\mathrm{min}$ & $>0.4$ & $>0.4$  & $>0.4$ & $>0.4$   \\ 
& $\met / \meff$     & $>0.3$     & $>0.25$ & $> 0.25$ & $>0.2$ \\ 
& $\meff$   & $>1000$  &  $>1000$ & $>500/1000$ & $>1100$ \\ 
\hline \hline
\end{tabular}
\caption{Event selection for each of the five overlapping signal regions ($\meff$, energy and momentum in GeV).
 Note that $\meff$ is defined with a variable number of jets, appropriate to each signal region. In the high mass selection, all jets with $\pt>40$~GeV are used to compute the $\meff$ value used in the final cut. The $\Delta \phi(\textrm{jet},\ptmiss)_\mathrm{min}$ cut is only applied up to the third leading jet. }
    \label{tab:0lep_SR}
  \end{center}
\end{table}

Dominant sources of background from Standard Model processes are  
W+jets, Z+jets, top pair, multi-jet and single top production. 
Non-collision background is found to be negligible. 

In order to estimate the backgrounds fully or partially in data-driven 
ways, five control regions (CRs) are defined for each of the five
signal regions (SRs), which are designed to provide uncorrelated data 
samples enriched with particular background sources. 
The observations in the CRs are used to estimate the background 
expectations in the SRs by using transfer factors, 
which are the ratios of expected event counts in the CR and SR
derived independently from the CR and SR themselves. 
The transfer factors are obtained from MC simulations 
for the W+jets, Z+jets, top backgrounds, whereas for the multijet
background, it is fully derived from data by smearing jets in 
the low \met ~events with jet response functions as is done
in~\cite{daCosta}.

For the Z+jets background, irreducible
contributions from Z($\rightarrow \nu \bar{\nu}$)+jets are dominant. 
Two CRs are considered that are enriched with $\gamma+$jets and 
Z($\rightarrow e^+ e^-, \mu^+ \mu^-$)+jets, 
which have similar 
kinematics. The momentum of the photon or the dilepton system from 
Z is added to the $\ptmiss$ to reproduce the \met ~distribution of the
Z($\rightarrow \nu \bar{\nu}$)+jets process. 

The W($\rightarrow l \nu$)+jets and top backgrounds are 
estimated from events containing one lepton 
and with \met $>$ 130 GeV. Furthermore, the 
transverse mass $M_T = \sqrt{2 p_T^l p_T^{\nu} (1 - \cos \Delta \phi(l,\nu))}$ 
is computed by assuming that the measured \met ~provides the neutrino information, and
is required to 
be between 30 GeV and 100 GeV to select W and top events. 
The events are then separated into two CRs 
using the b-tagging information, where jets arising from b-quarks
are identified using the impact parameter and secondary
vertex information. 
The lepton in the events is treated as a jet when computing 
the kinematic variables, since the dominant sources of W+jets 
and top backgrounds contain W's decaying into hadronic taus and 
tau-neutrinos.

Finally, for the multijet background, the control region is 
chosen such that the cut on $\Delta \phi(\textrm{jet},\ptmiss)_\mathrm{min}$
is reversed and tightened to 
$\Delta \phi(\textrm{jet},\ptmiss)_\mathrm{min} < 0.2$.
In such events, $\ptmiss$ is aligned with jets in the 
transverse plane, and \met ~originates from the mismeasurement of jets
or neutrino emission from heavy flavor decays within jets. 
An additional and separate CR was considered to estimate the 
impact of the dead region in the LAr EM barrel calorimeter. 

For the transfer factors derived from the MC simulation, important sources of 
systematic uncertainties are the jet energy scale, jet energy 
resolution, MC modeling uncertainties such as renormalization and 
factorization scale and parton distribution function (PDF) uncertainties, 
and uncertainty arising from the presence of pile-up. 
Additional uncertainties originate from photon and lepton trigger and
reconstruction efficiencies, photon and lepton energy scale and 
resolution (for estimating the W+jets, Z+jets, and top backgrounds),  
b-tag/veto efficiency (for W+jets and top), photon acceptance 
and background (for Z+jets), and MC statistics. For the data-driven
estimation of the multijet background, the main source of uncertainties
for the transfer factor is the modeling of the non-Gaussian tail
of the jet response function. 


\begin{table*}[t]
\small
\begin{center}
\begin{tabular}{|c|c|c|c|c|c|}
\hline\hline
\multirow{2}{*}{Process}     &\multicolumn{5}{|c|}{Signal Region} \\ 
\cline{2-6} & \multirow{2}{*}{$\ge 2$-jet} & \multirow{2}{*}{$\ge 3$-jet} & $\ge 4$-jet,  & $\ge 4$-jet,  & \multirow{2}{*}{High mass} \\ 
                                           & & & $\meff>500$~GeV & $\meff>1000$~GeV & \\ \hline \hline
$Z/\gamma$+jets  & $ 32.3\pm\ph2.6\pm\ph6.9$ & $ 25.5\pm\ph2.6\pm\ph4.9$ & $ 209\pm\ph9\pm\ph38$ & $ 16.2\pm\ph2.2\pm\ph3.7$ & $ \ph3.3\pm\ph1.0\pm\ph1.3$ \\
 $W$+jets &$ 26.4\pm\ph4.0\pm\ph6.7$ &$ 22.6\pm\ph3.5\pm\ph5.6$ &$ 349\pm30\pm122$ &$ 13.0\pm\ph2.2\pm\ph4.7$ &$ \ph2.1\pm\ph0.8\pm\ph1.1$ \\
$t\bar{t}$+ single top  &$ \ph3.4\pm\ph1.6\pm\ph1.6$ &$ \ph5.9\pm\ph2.0\pm\ph2.2$ &$ 425\pm39\pm\ph84$ &$ \ph4.0\pm\ph1.3\pm\ph2.0$ &$ \ph5.7\pm\ph1.8\pm\ph1.9$ \\
QCD multi-jet   &$ 0.22\pm0.06\pm0.24$ &$ 0.92\pm0.12\pm0.46$ &$ \ph34\pm\ph2\pm\ph29$ &$ 0.73\pm0.14\pm0.50$ &$ 2.10\pm0.37\pm0.82$ \\

\hline

Total  &$ 62.4\pm\ph4.4\pm\ph9.3$ &$ \ph54.9\pm\ph3.9\pm\ph7.1$ &$ 1015\pm41\pm144$ &$ 33.9\pm\ph2.9\pm\ph6.2$ &$ 13.1\pm\ph1.9\pm\ph2.5$ \\

\hline\hline

Data  & 58 & 59 & 1118 & 40 & 18 \\

  \hline\hline
 \end{tabular}
\caption{\label{tab:SRevc} Fitted background components in each SR, compared with the observation. 
In each case the first (second) quoted uncertainty is statistical (systematic). Estimates of background 
components are partially correlated and hence the uncertainties (statistical and systematic) on the total 
background estimates do not equal the quadrature sums of the uncertainties on the components. }
 \end{center}
 \end{table*}

Table~\ref{tab:SRevc} shows the numbers of Standard Model background
events expected in the signal regions, and the numbers of observed
events. The \meff ~distributions are shown in Figure~\ref{fig:0lep_meff} for 
each signal region. No excess is observed, and thus limits are obtained 
for each channel. The $\text{CL}_\text{s}$ prescription ~\cite{cls}
is used to set the 
exclusions. 
Figure~\ref{fig:0lep_limits} 
shows the combined exclusion limits 
for the simplified models with $m_{\text{LSP}} = 0$. 


\begin{figure} [p]
\centering
\begin{tabular}{cc}
\includegraphics[width=0.4\linewidth]{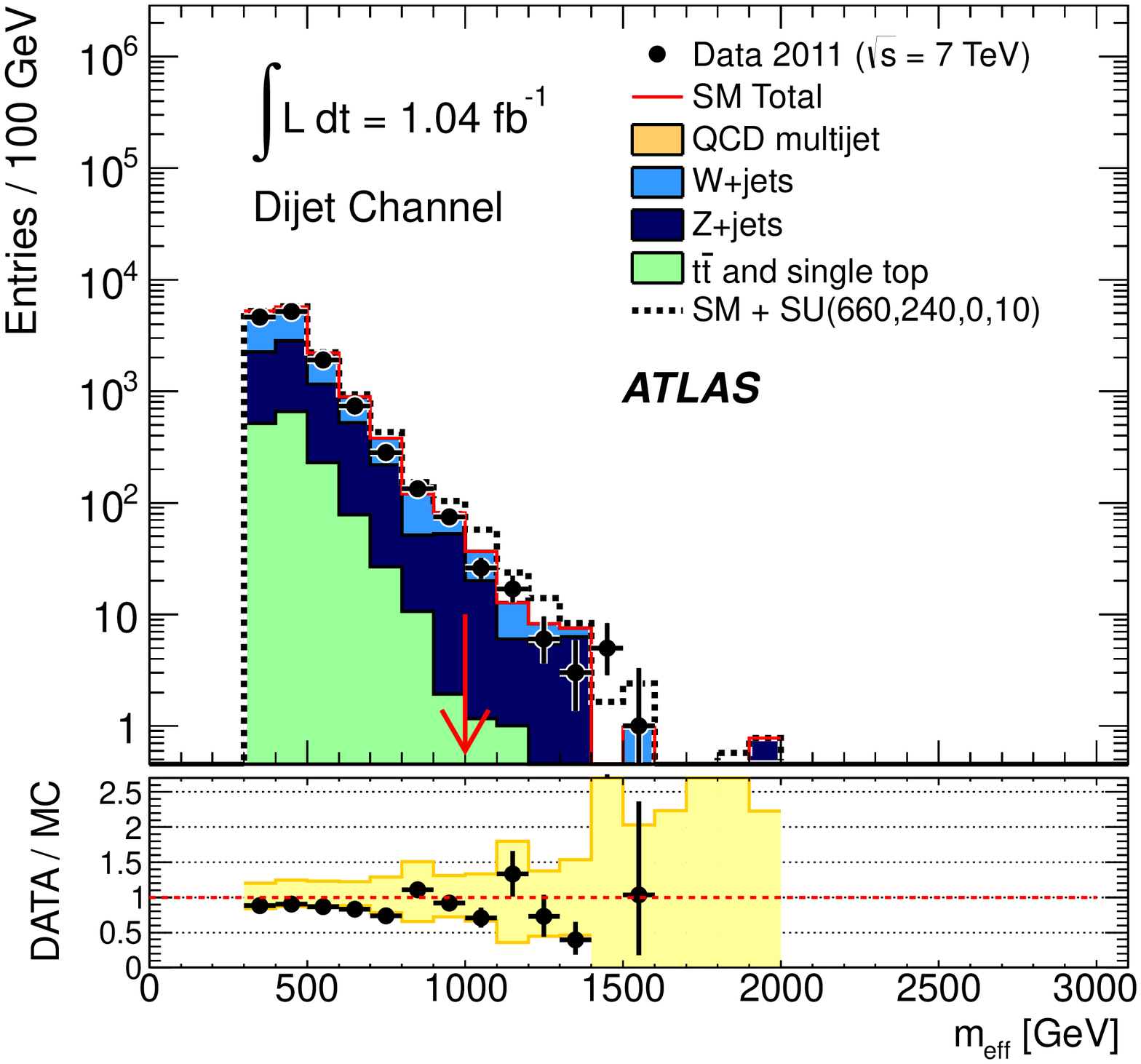}
\includegraphics[width=0.4\linewidth]{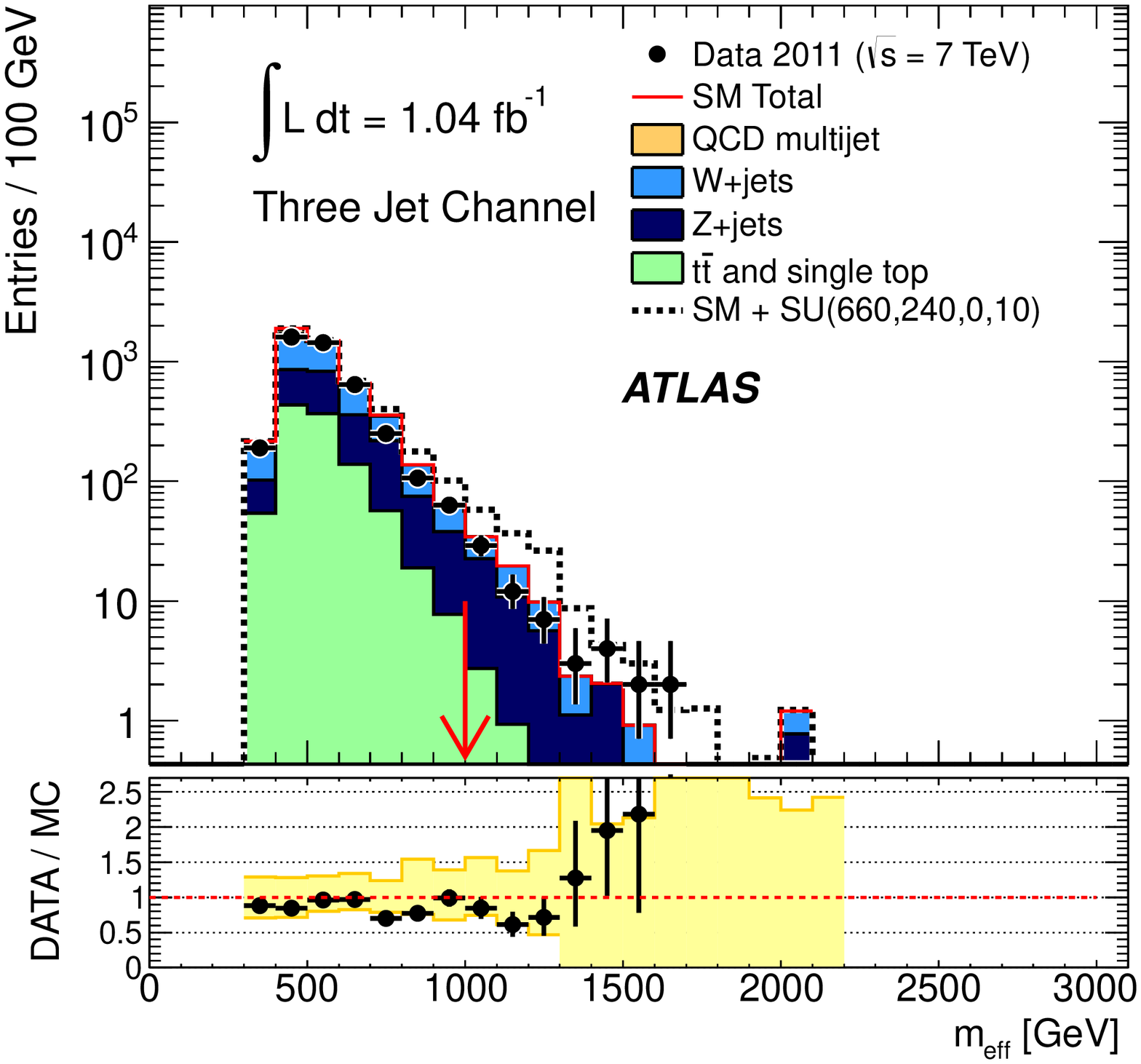} \\
\includegraphics[width=0.4\linewidth]{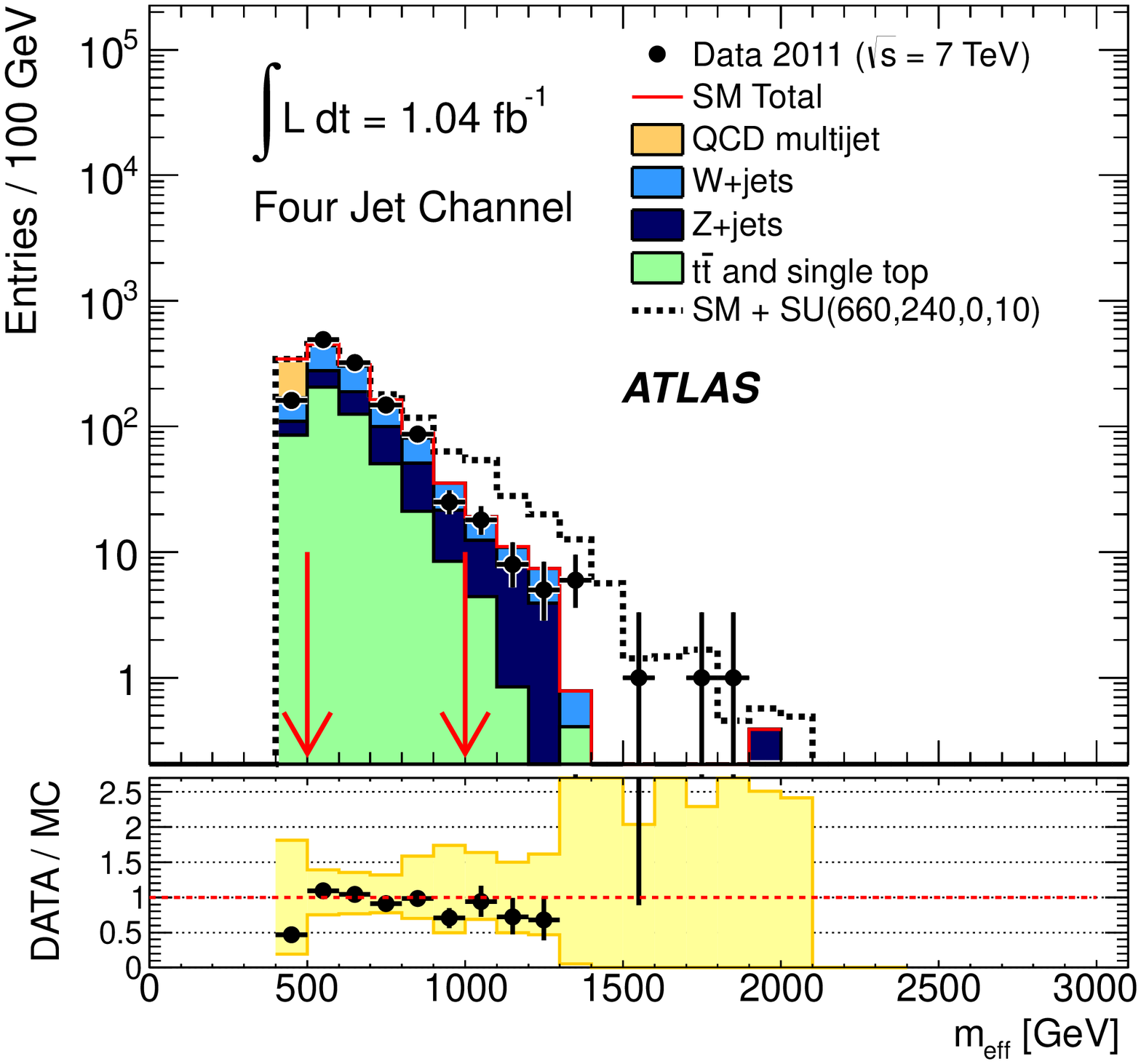}
\includegraphics[width=0.4\linewidth]{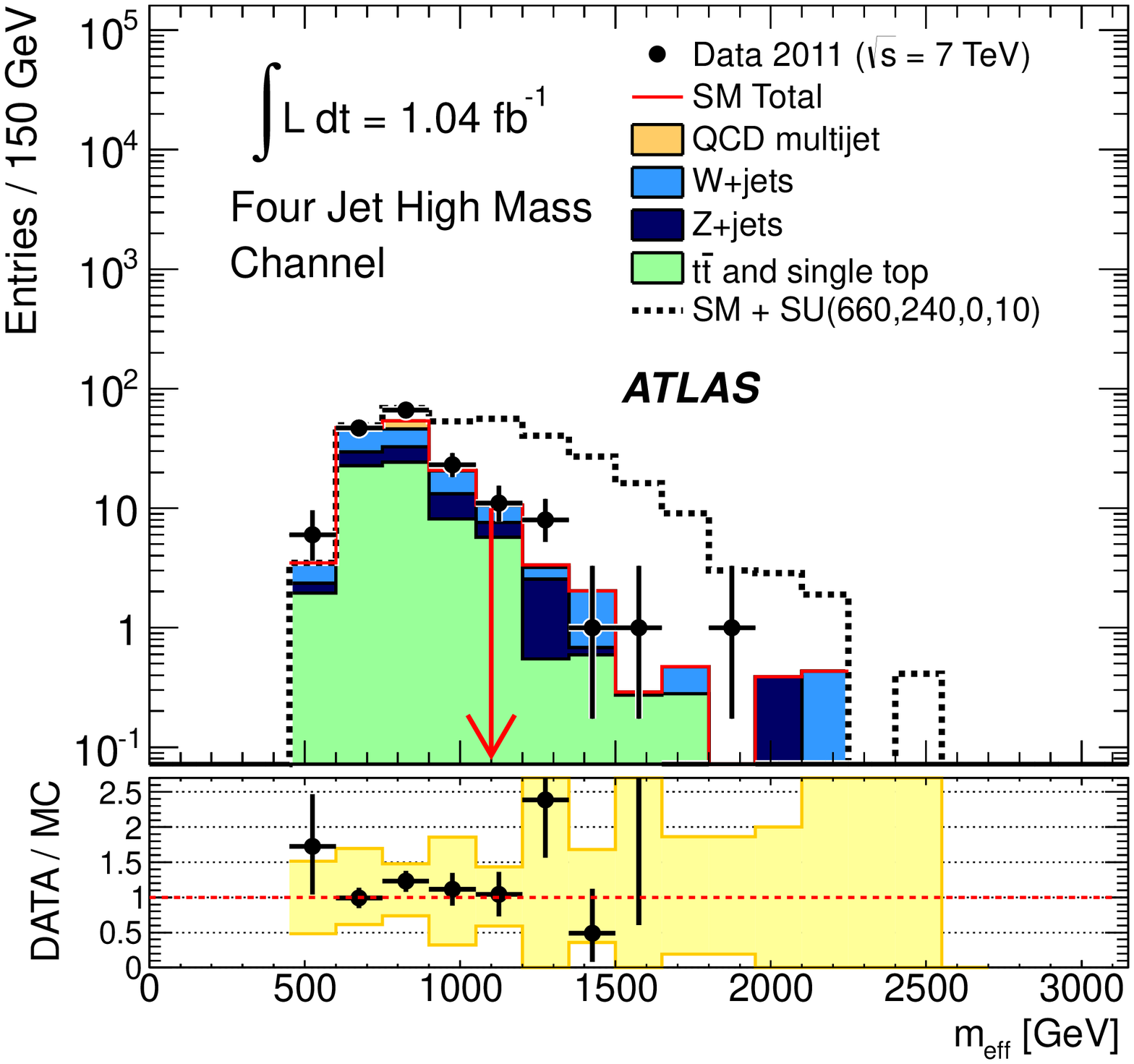} 
\end{tabular}
\caption{The \meff ~distributions for the dijet (top left), three jet (top
right), four jet (bottom left), and four jet high mass (bottom right) 
channels.}
\label{fig:0lep_meff}
\end{figure}

\begin{figure}[p]
\centering
\includegraphics[width=80mm]{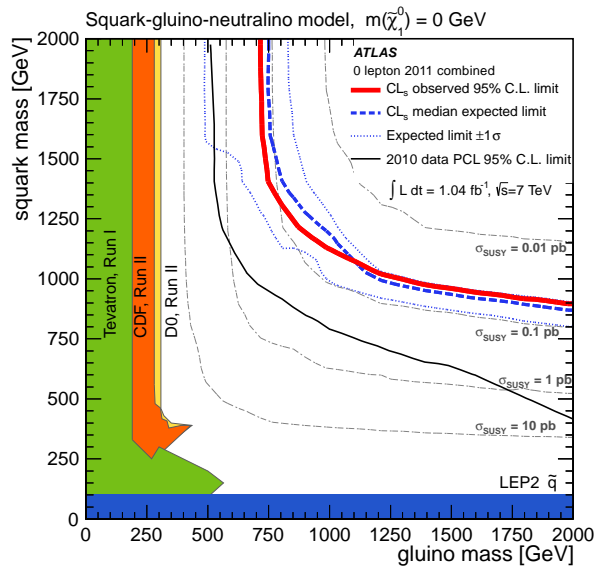}
\caption{Combined exclusion limit for the simplified models with 
$m_{\text{LSP}} = 0$. The red bold line shows the observed limit at 95\%
CL, and the blue dashed line corresponds to the median expected 95\% CL 
limit. The dotted blue lines show the expected 68\% and 99\% CL expected
limits. } 
\label{fig:0lep_limits}
\end{figure}

\section{Same-sign Dilepton Search}

A search in the same-sign dilepton channel is considered
to be a clear channel for the supersymmetry search 
due to very low expected background
from the Standard Model~\cite{barnett}. 

Here, simplified models are considered for the same-sign 
same-flavor squark pair production contributing to the 
signal region as shown in 
Figure~\ref{fig:simplified_sqsq}~\cite{okawa}\cite{simpleSS}. 
As such a process 
originates from the same flavor quark interactions, the 
contributions from the third generation squarks can be
disentangled~\cite{alwall}. For these simplified models,
there are three mass parameters:  
squark, weakino ($\chi_1^{\pm}, \chi_2^0$), and the LSP. 
Branching ratios are also free parameters, and the ones
relevant for the diagrams are mentioned below. 
\begin{equation}
\label{eq:br_ww}
Br(\tilde{q}\tilde{q} \rightarrow qq \ell^{\pm}\nu \ell^{\pm}\nu \tilde{\chi}^0_1 \tilde{\chi}^0_1) = \left[ Br(\tilde{q} \rightarrow q\tilde{\chi}_1^{\pm}) Br(\tilde{\chi}_1^{\pm} \rightarrow \ell^{\pm}\nu \tilde{\chi}^0_1) \right]^2
\end{equation}
\begin{equation}
\label{eq:br_wz}
Br(\tilde{q}\tilde{q} \rightarrow qq \ell\nu \ell^+\ell^- \tilde{\chi}^0_1 \tilde{\chi}^0_1) = 2 Br(\tilde{q} \rightarrow q\tilde{\chi}_1^{\pm}) Br(\tilde{q} \rightarrow q\tilde{\chi}^0_2) Br(\tilde{\chi}_1^{\pm} \rightarrow \ell\nu \tilde{\chi}^0_1) Br(\tilde{\chi}^0_2 \rightarrow \ell^+\ell^- \tilde{\chi}^0_1) 
\end{equation}

\begin{figure} [t]
\centering
\begin{tabular}{cc}  
\includegraphics[width=0.4\linewidth]{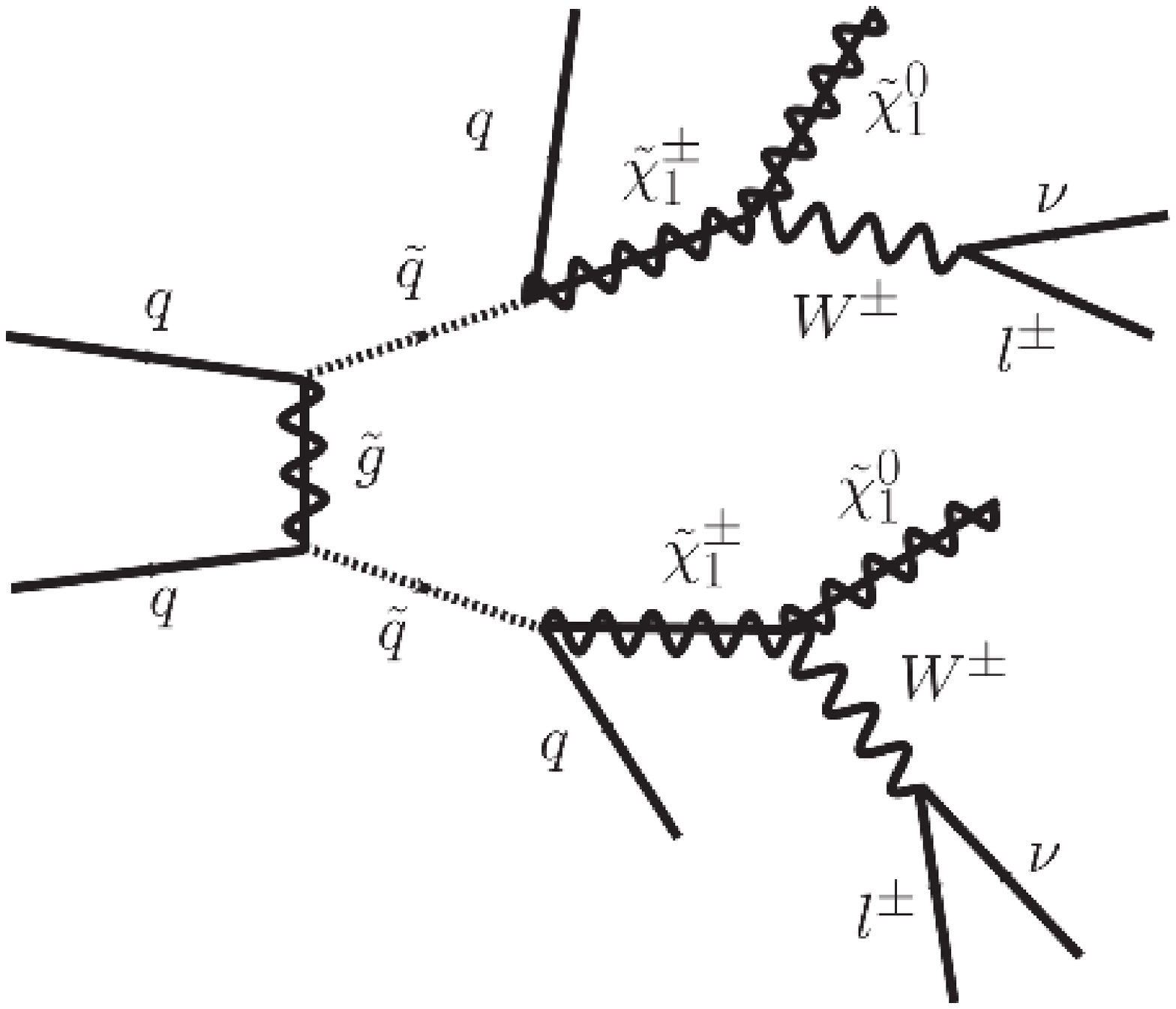}
\includegraphics[width=0.4\linewidth]{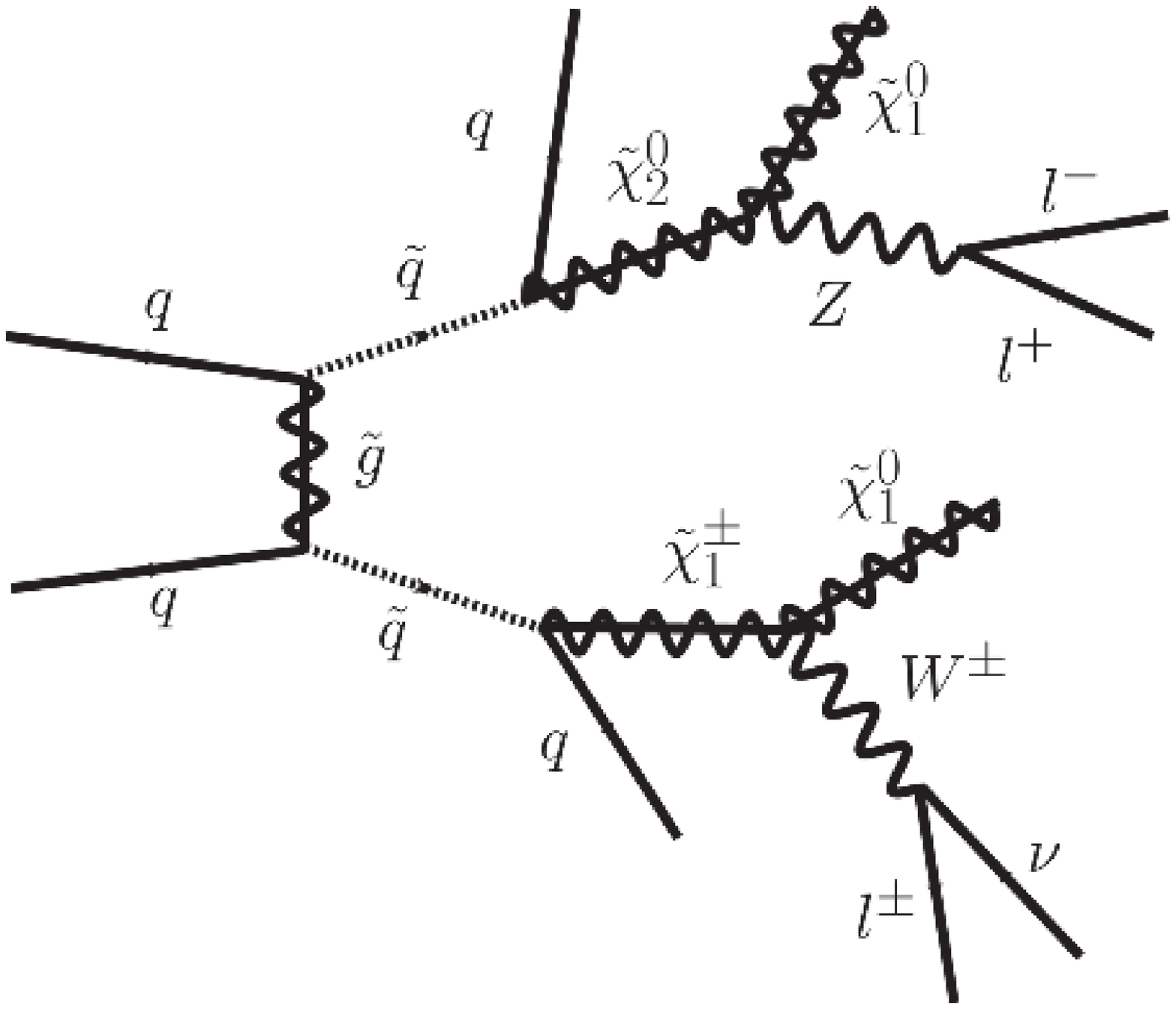}
\end{tabular}
\caption{Simplified models with dilepton (left) and trilepton (right) final 
states considered for the same-sign dilepton search.}
\label{fig:simplified_sqsq}
\end{figure}

For this channel,
events are required to have at least one reconstructed primary vertex with 
at least five associated tracks. 
Furthermore, the distance between the $z$ coordinate of the primary vertex 
and that of the extrapolated muon track at the point closest to the primary 
vertex must be less than 10 mm to suppress cosmic background. 
The signal region is defined as events with exactly two identified 
same-sign leptons (electrons and muons)
with $p_T>20\GeV$ and $\met > 100$ GeV.

Backgrounds from several Standard Model processes could contaminate 
the signal regions. The main background to the same-sign dilepton final 
state arises from $W$+jets and QCD multijet production, where one or 
more jets are misidentified as isolated leptons, 
which is referred to as ``fake lepton'' background. 
The other significant backgrounds arise from charge misidentification 
of an electron due to a hard bremsstrahlung process in the $t\bar t$ process. 
Contributions from $Z$+jets are negligible due to a high \met~cut. 
Contributions from the diboson production are also estimated.
The fake background is estimated in a data-driven way using the matrix 
method~\cite{2lep-paper}, 
and MC-based estimations are used for the other backgrounds. 
The cosmic background was considered and found to be negligible. 

The systematic uncertainties on the data-driven fake background 
estimates mainly come from the parametrization of the fake rate. For 
the MC-driven background estimates, 
the jet energy scale and resolution, lepton energy scale, resolution and 
identification, luminosity, cross section, and parton distribution function 
(PDF) uncertainties are considered. More details are mentioned 
in~\cite{2lep-paper}.

\begin{figure}[t]
\centering
\includegraphics[width=80mm]{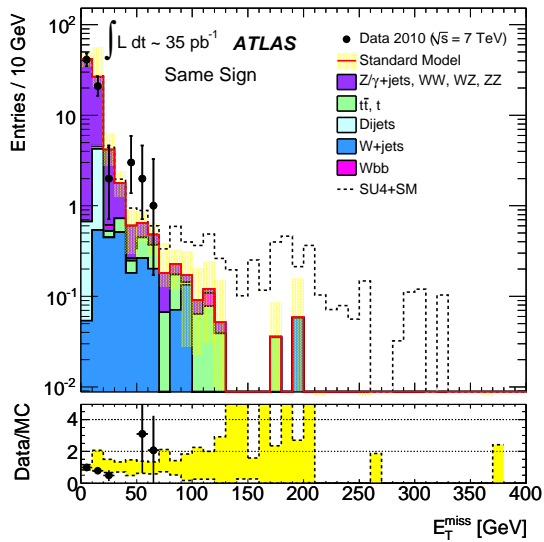}
\caption{Distribution of \met ~in the same-sign dilepton channel.} 
\label{fig:ss2lep_etmiss}
\end{figure}

\begin{table}[t] \small
  \begin{center}
   \caption{Summary of the background yields and observed number of events for the same sign dilepton channels.}
  \begin{tabular}{lccc} \hline \hline
    & $e^{\pm}e^{\pm}$ &$e^{\pm}\mu^{\pm}$ & $\mu^{\pm}\mu^{\pm}$
    \\\hline 
  Fakes              & 0.12 $\pm$ 0.13  &  0.030 $\pm$ 0.026 & 0.014 $\pm$ 0.010 \\
  Di-bosons          & 0.015 $\pm$ 0.008 & 0.035 $\pm$ 0.014  & 0.021 $\pm$ 0.006 \\
  Charge misidentification        & 0.019 $\pm$ 0.008 & 0.026 $\pm$ 0.011  & - \\ 
  Cosmics            & -     & $0^{+1.17}_{-0}$ & - \\  \hline
  Total              & 0.15 $\pm$ 0.13 & 0.09$^{+1.17}_{-0.03}$ & 0.04 $\pm$ 0.01 \\ \hline
  Data               & 0 &  0 &  0 \\
\hline
\hline
\end{tabular}
    \label{tab:expectedEvents}
  \end{center}
\end{table}

No excess above the Standard Model prediction is observed as shown in 
Figure~\ref{fig:ss2lep_etmiss} and Table~\ref{tab:expectedEvents}, and
upper limits on the cross-section times branching ratios for new physics 
based on the simplified models are obtained~\cite{simpleSS}. 
Classical confidence intervals in the theoretical cross section are 
constructed by generating ensembles of pseudo-experiments that describe 
the expected fluctuations 
of statistical and systematic uncertainties on both signal and backgrounds, 
following the likelihood ratio ordering prescription proposed by Feldman 
and Cousins~\cite{FC}. The PDF uncertainty for the signals is estimated
in the same way as the SM background, adopting a conservative
uncertainty of 5.5\%. 

Figure~\ref{fig:limit2D} shows 95\% confidence level (CL) observed upper limits for 
each diagram as a function of $m_{\tilde{q}}$ and 
$m_{\tilde{\chi}_1^{\pm},\tilde{\chi}_2^0}$ for the fixed LSP mass of 50 GeV. 
Each figure corresponds to a specific diagram, where the
branching ratios are explicitly described in 
(\ref{eq:br_ww})-(\ref{eq:br_wz}). Corresponding limit plots for the 
fixed LSP masses of 100 and 200 GeV, and for the diagram from a four-lepton 
final state are described in~\cite{simpleSS}.

For each diagram, a grid of 26 signal points is produced and used for the 
limit setting.
The upper limits are interpolated linearly from those points in 
three coordinates. The impact of the interpolation on the signal acceptance 
is found to be below a few percent and therefore does not impact the results.


\begin{figure} [t]
\centering
\begin{tabular}{cc}  
\includegraphics[width=0.4\linewidth]{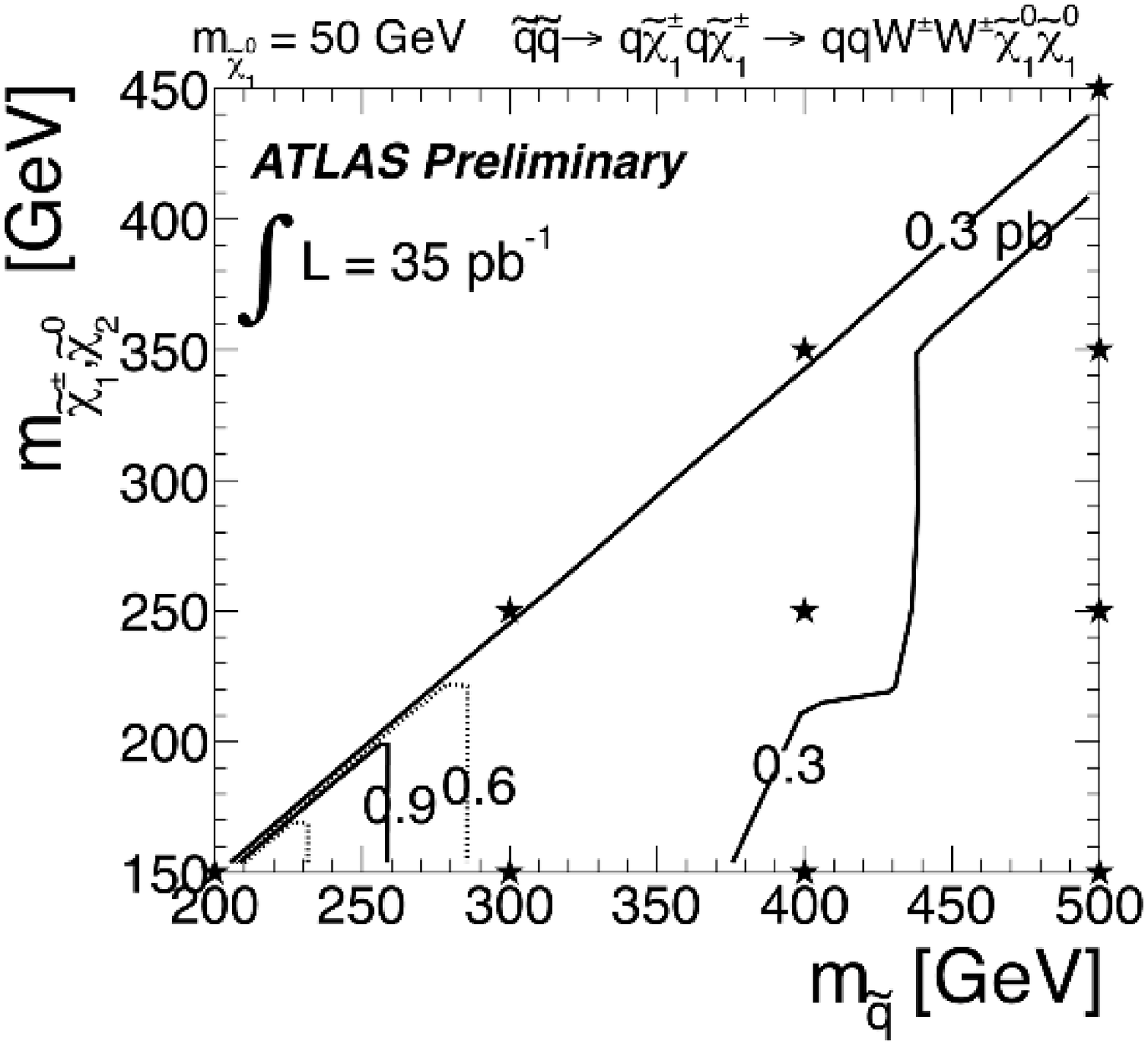}
\includegraphics[width=0.4\linewidth]{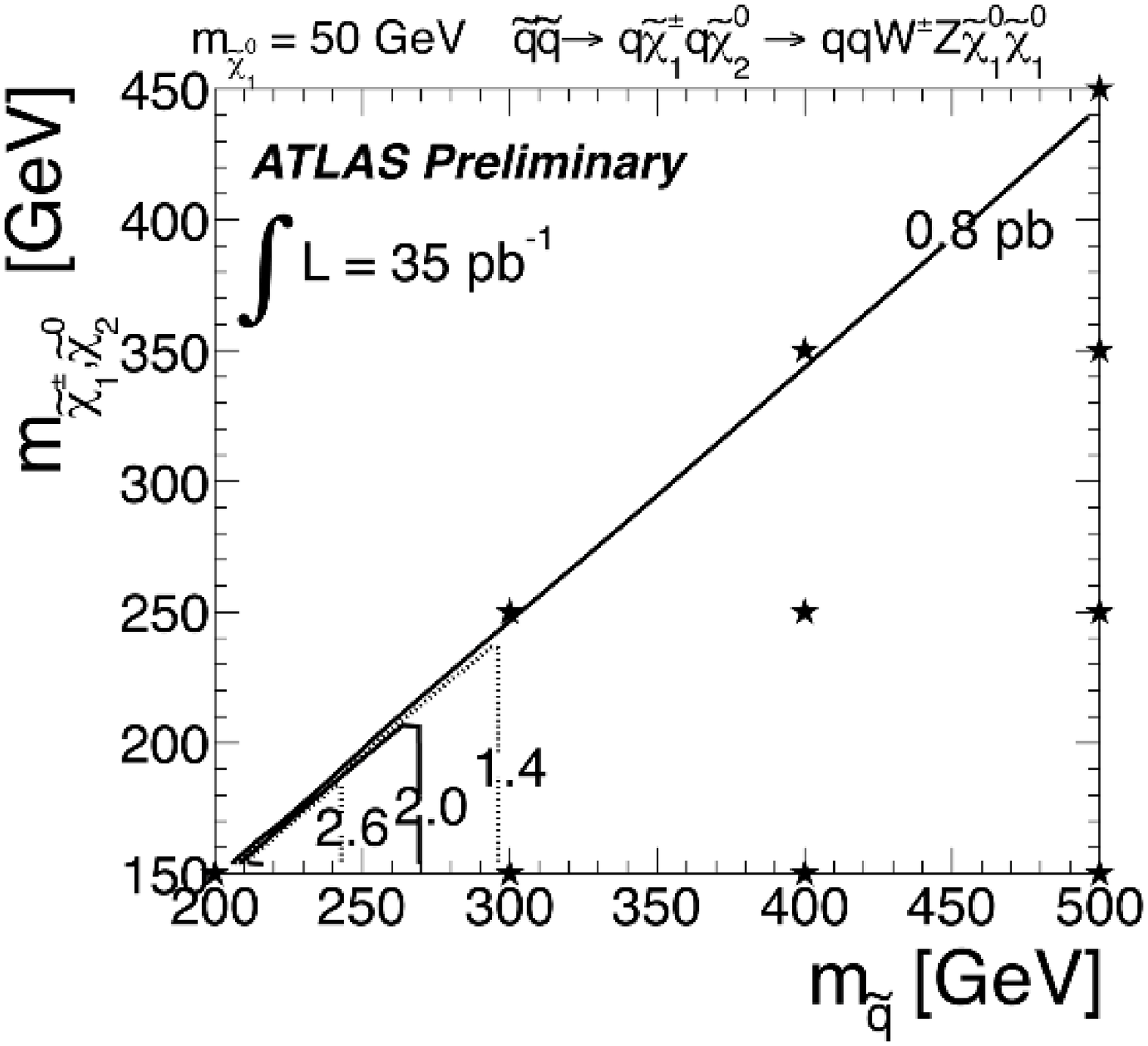}
\end{tabular}
\caption{Observed upper limits (95\% CL) on $\sigma \times$ Br in pb for the
$\tilde{q}\tilde{q}\rightarrow q\tilde{\chi}_1^{\pm}q\tilde{\chi}_1^{\pm} \rightarrow qqW^{\pm}W^{\pm}\tilde{\chi}^{0}_{1}\tilde{\chi}^{0}_{1}$ process (left) 
and $\tilde{q}\tilde{q}\rightarrow q\tilde{\chi}_1^{\pm}q\tilde{\chi}_2^0 \rightarrow qqW^{\pm}Z\tilde{\chi}^{0}_{1}\tilde{\chi}^{0}_{1}$ (right) as 
a function of $m_{\tilde{q}}$ and $m_{\tilde{\chi}_1^{\pm}, \tilde{\chi}_2^0}$ for the fixed LSP mass of 50 GeV.}
\label{fig:limit2D}
\end{figure}

\section{Conclusions}

No excess is found in the no-lepton channel with 1.04 \ifb of data from 
2011, and in the same-sign dilepton channel using 2010 data of 35 \ipb.  
The results are interpreted with simplified models. 
They allow to scan wide range of sparticle mass planes, and 
to understand which phase space is currently covered. 
The simplified
model approach provides model-independent interpretations and will also
serve as an interface to map the results to specific models even beyond
SUSY models. 
Interplay between different search channels is gaining more interest
and importance. 
More simplified model results are currently under investigation.

\begin{acknowledgments}

The author would like to thank NSF and the DPF 2011 organizing committee for the partial 
financial support.  

\end{acknowledgments}

\bigskip 

\bibliographystyle{atlasnote}
\bibliography{dpf2011_template}

\providecommand{\href}[2]{#2}\begingroup\raggedright\begin{thebibliography}{10}

\bibitem{ATLAS_exp}
{ATLAS Collaboration,} JINST {\bf 3} (2008)  S08003.

\bibitem{LHC}
{L. Evans and P. Bryant (editors),} JINST {\bf 3} (2008)  S08001.

\bibitem{pub_lumi}
{ATLAS Collaboration,}
  \url{https://twiki.cern.ch/twiki/bin/view/AtlasPublic/LuminosityPublicResult%
s}  .

\bibitem{marmo}
{N. Arkani-Hamed et al.,} arXiv:hep-ph/0703088 (2007)  .

\bibitem{alwall}
{J. Alwall et al., Phys. Rev. D {\bf 79} (2000) 015005; J. Alwall et al., Phys.
  Rev. D {\bf 79} (2000) 075020}.

\bibitem{alves}
{D. Alves et al., Phys. Lett. B \textbf{702} (2011) 64; D. Alves et al.,
  arXiv:1102.5338 (2011); D. Alves et al., arXiv:1105.2838 (2011)}.

\bibitem{lhcnewphys}
{Signatures of New Physics at the LHC,} \url{http://lhcnewphysics.org}  .

\bibitem{bart}
{B. Butler}, in these proceedings.

\bibitem{msugra}
{A. H. Chamseddine et al., Phys. Rev. Lett. \textbf{49} (1982) 970; R. Barbieri
  et al., Phys. Lett. B \textbf{119} (1982) 343; L. E. Ibanez, Phys. Lett. B
  \textbf{118} (1982) 73; L. J. Hall et al., Phys. Rev. D \textbf{27} (1983)
  2359; N. Ohta, Prog. Theor. Phys. \textbf{70} (1983) 542}.

\bibitem{cmssm}
{G.~L.~Kane et al., Phys. Rev. D {\bf 49} (1994) 6123}.

\bibitem{rizzo}
{C.F. Berger et al., JHEP 0902:023 (2009); J. Conley et al., Eur. Phys. J. C
  \textbf{71} (2011) 1697; J. Conley et al., arXiv:1103.1697 (2011); }.

\bibitem{UED}
{C. Macesanu et al.,} Phys. Lett. B {\bf 546} (2002)  253.

\bibitem{lumi2011}
{ATLAS Collaboration,} ATLAS-CONF-2011-116  (2011),
  \url{http://cdsweb.cern.ch/record/1376384}.

\bibitem{lumi}
{ATLAS Collaboration,} EPJC {\bf 71} (2011)  1630.

\bibitem{2lep-paper}
{ATLAS Collaboration,} EPJC {\bf 71} (2011)  1682.

\bibitem{antiKt}
{M.~Cacciari and G.~P.~Salam,} JHEP {\bf 0804} (2008)  063.

\bibitem{antiKt2}
{M.~Cacciari and G.~P.~Salam,} Phys. Lett. B {\bf 641} (2006)  57.

\bibitem{calocluster}
{ATLAS Collaboration,} ATL-LARG-PUB-2008-002  (2008),
  \url{http://cdsweb.cern.ch/record/1099735}.

\bibitem{atlas-jetcal}
{ATLAS Collaboration,} ATLAS-CONF-2011-032  (2011),
  \url{http://cdsweb.cern.ch/record/1337782}.

\bibitem{atlas-electron2011}
{ATLAS Collaboration,} ATL-PHYS-PUB-2011-006  (2011),
  \url{http://cdsweb.cern.ch/record/1345327}.

\bibitem{atlas-muon2011}
{ATLAS Collaboration,} ATLAS-CONF-2011-063  (2011),
  \url{http://cdsweb.cern.ch/record/1345743}.

\bibitem{ATLAS_detPerf}
{ATLAS Collaboration,} arXiv:0901.0512 (2008)  .

\bibitem{jetcleaning}
{ATLAS Collaboration,} ATLAS-CONF-2010-038  (2010),
  \url{http://cdsweb.cern.ch/record/1277678}.

\bibitem{Meff}
{D. Tovey,} EPJ Direct {\bf 4} (2002)  N4.

\bibitem{daCosta}
{ATLAS Collaboration,} arXiv:1109.6572  (2011), submitted to Phys. Lett. B.

\bibitem{cls}
{A. Read,} Journal of Physics G: Nucl. Part. Phys. {\bf 28} (2002)  2693.

\bibitem{barnett}
{R. M. Barnett et al., Phys. Lett. B \textbf{315} (1993) 349; R. M. Barnett et
  al., Proc. of 1988 Summer Study on High Energy Physics in the 1990's,
  Snowmass, Colorado, 1988.}

\bibitem{okawa}
H.~Okawa, presentation at Characterization of new physics at the LHC II, CERN,
  November 6, 2010.

\bibitem{simpleSS}
{ATLAS Collaboration,} ATLAS-CONF-2011-091  (2011),
  \url{http://cdsweb.cern.ch/record/1360190}.

\bibitem{FC}
{G. J. Feldman and R. D. Cousins,} Phys. Rev. D {\bf 57} (1998)  3873.

\end{thebibliography}\endgroup


\begin{thebibliography}{9}   

\bibitem{lhcnewphys} http://www.lhcnewphysics.org/

\bibitem{rizzo} J. Conley et al., arXiv:1103.1697 (2011)

\bibitem{alwall} J. Alwall et al., Phys. Rev. D \textbf{79}, 075020, 38 (2009)

\bibitem{charm07}   http://www.lepp.cornell.edu/charm07/

\bibitem{templates-ref} http://www.slac.stanford.edu/econf/editors/eprint-template/instructions.html

\end{thebibliography}

\end{document}